\documentclass[a4paper,11pt]{article}
\pdfoutput=1 

\usepackage{jcappub} 

\usepackage[T1]{fontenc} 
\usepackage{fontawesome5}
\usepackage{float}
\usepackage{soul}

\usepackage{amsmath}

\DeclareMathOperator\erfc{erfc}
\newcommand{\bh}{\rm PBH}

\title{\boldmath Constraints on primordial black holes for nonstandard cosmologies}

\author[a]{Tadeo D. Gomez-Aguilar,}
\author[a]{Luis E. Padilla,\note{Corresponding author.}}
\author[b]{Encieh Erfani,}
\author[a,1]{Juan Carlos Hidalgo}

\affiliation[a]{Instituto de Ciencias Físicas, Universidad Nacional Autónoma de México,\\ 62210, Cuernavaca, Morelos, México.}
\affiliation[b]{PRISMA+ Cluster of Excellence \& Mainz Institute for Theoretical Physics,\\ 
Johannes Gutenberg-Universität Mainz, 55099 Mainz, Germany.}

\emailAdd{tadeo.dga@icf.unam.mx}
\emailAdd{lepadilla@icf.unam.mx}
\emailAdd{erfanien@uni-mainz.de}
\emailAdd{hidalgo@icf.unam.mx}

\abstract{We study how the bounds on the abundance of Primordial Black Holes (PBHs) and the constraints on power spectrum are modified if a non-standard evolution phase takes place between the end of inflation and the Standard radiation-dominated (RD) universe after inflation. The constraints on PBH abundance and power spectrum are computed using the new, freely available, \href{https://github.com/TadeoDGAguilar/PBHBeta}{\faGithubSquare}~\texttt{PBHBeta} library, which accounts for the effects of non-standard expansion and specific criteria for PBH formation in such non-standard scenarios. As working examples, we consider three different cases: a pure matter-dominated (MD) phase, a scalar field-dominated ($\varphi$D) universe, and a stiff fluid-dominated (SD) scenario.
While the background expansion is the same for the MD and $\varphi$D scenarios, the PBH formation criteria lead to different constraints to power spectrum. On the other hand, the duration of the non-standard expansion phase alters the bounds, with longer MD periods resulting in weaker constraints on power spectrum, and longer SD scenarios leading to an enhanced abundance due to the dust-like redshifting of PBHs. The modifications to the constraints are reported in all cases and we highlight those where the power spectrum may be significantly constrained.}

\begin{document}
\maketitle
\flushbottom

\section{Introduction}

{More than fifty years have passed since the first studies on Primordial Black Holes (PBHs) \citep{zel1967hypothesis}. Although there is still no definitive evidence of their existence, recent advances in gravitational wave astronomy have increased the interest in studying these objects. One of the reasons for such interest is the wide mass spectrum, which can accommodate observed phenomena with no definite explanation. A characteristic observable signature of PBHs comes from Hawking radiation \citep{Hawking:1974rv}. According to the theory, PBHs with masses $M_{\rm PBH}\lesssim 10^{15} \rm{g}$ should have been evaporated. However, with adequate numbers, these light PBHs can explain different astrophysical observations, as the extragalactic \citep{Page:1976wx} and galactic \citep{refId0} $\gamma$-ray backgrounds or some short-period $\gamma$-ray bursts \citep{Cline_1997}. On the other hand, PBHs with a mass larger than $10^{15} \rm{g}$ can contribute to part of the Dark Matter. Even if they do not contribute significantly to the dark matter in a certain part of the spectrum, their existence could explain observations such as the seeds of supermassive black holes \citep{PhysRevD.66.063505} or observations of black holes in the intermediate mass range \citep{sasaki2016primordial}.}

{In the standard cosmology\footnote{We will throughout refer to the scenario where the universe is radiation-dominated immediately after inflation as the standard cosmology or the Standard Big Bang cosmology.}, in which the universe is radiation-dominated (RD) after inflation, the threshold density contrast $\delta_{\rm c}$ for the formation of PBHs is $0.41$ \citep{Harada:2013epa}.}
{ However, in alternative models, PBHs could also form following different criteria. For instance, if the equation of state of the universe at early times departs from the RD, i.e. having a matter-dominated (MD) era after inflation. Some examples that would allow this dust-like evolution in the early universe are: a slow post-inflationary reheating \citep{PhysRevLett.73.3195,PhysRevD.56.3258,Allahverdi:2010xz,2019arXiv190704402L,RevModPhys.78.537,Amin2014eta}, and massive metastable particles dominating the universe after the inflationary epoch \citep{Nanopoulos:1979gx, Khlopov:1980mg}. Particular attention must be payed to the recurrent picture of scalar fields dominating the post-inflationary universe, and we will explore this case in detail in section \ref{Sec2}. All these alternatives to the standard Big Bang scenario alter the expansion rate of the universe and can bring important consequences for the formation and abundance of PBHs, as we shall review in this paper.} 

Current observations of the temperature anisotropies in the Cosmic Microwave Background (CMB) limit the energy scale during inflation to be smaller than $\mathcal{O}(10^{16})~\rm{GeV}$ (see e.g. \citep{Planck:2018jri}). On the other hand, observations of Big Bang Nucleosynthesis (BBN) require that the universe become RD at the energy scale $1-100~\rm{MeV}$ \citep{1990eaun.book.....K, Tytler:2000qf,deSalas2015glj,Hasegawa2019jsa}. The expansion history between these two epochs is not well constrained by observations and, in principle, a nonstandard evolution could occur during this period. 

Many of the above and other models alternative to the standard Big Bang are motivated by the current questions of cosmology. In the present day, the nature of inflation, dark matter, and dark energy are largely unknown. In particular, the precise transition between inflation and the radiation domination era is poorly understood \cite{Amin2014eta, Allahverdi:2010xz, 2021arXiv210410552E}, in part due to the scarce observational probes associated with such a period. It is only through high-frequency gravitational waves \citep{Giovannini:1998bp, Giovannini:1999bh, Domenech:2023jve} or cosmological effects of the Primordial Black Hole spectrum \citep{Padilla:2023lbv} that one can address the properties of such transition.  In addition, some tensions have become increasingly evident as observational data become more precise, such as the Hubble tension \citep{2019NatAs...3..891V}, the coincidence problem \citep{PhysRevLett.82.896}, and the small scale structure problem \citep{doi:10.1146/annurev-astro-091916-055313}. These inconsistencies and uncertainties are eased with alternatives to the standard Big Bang cosmology that could, in some way, help to better describe our universe.

The main idea of the present work is to test alternatives to the period between the inflationary and the BBN stages. We are interested in cosmological models that deviate from the standard Big Bang and how they influence the current constraints on the power spectrum from PBHs abundance bounds. Some possibilities of nonstandard Big Bang come from theories like supersymmetry \citep{PhysRevD261231, Dine:1995uk, Moroi:1999zb,Lyth:1995hj}, string theory  \citep{Cicoli:2023opf, Asaka:1999xd, Dutta:2016htz}, and modified gravity \citep{TDamour_1992, Cline:1999ts, Csaki:1999jh}. Previous studies have investigated particular expansion histories, addressing the formation of PBHs and their implications for the power spectrum \citep{Josan:2009qn, Cole:2017gle, Carr:2017edp, Bhattacharya:2021wnk, Bhattacharya:2019bvk}. In contrast, the present study takes a more general perspective on the conditions to form PBHs and how these evolve within non-standard scenarios, focusing on the modification of the constraints to the PBH abundance\footnote{Throughout this paper, we consider that the PBHs formed are of the Schwarzschild kind, an assumption justified below.}. {For example, we incorporate general criteria that may modulate the conditions for PBH formation such as the sphericity criterion \citep{Harada:2016mhb}, and the conservation of angular momentum \citep{Harada:2017fjm}. {We include also the influence of inhomogeneities within the configuration  in the formation of PBHs \cite{Khlopov:1980mg,Kokubu:2018fxy,Passaglia:2021jla,Green:2020jor,ShamsEsHaghi:2022azq,Harada:2022xjp}}.}

The main idea of the present work is to test alternatives to the period between the inflationary and the BBN stages. We are interested in cosmological models that deviate from the standard Big Bang and how they influence the current constraints on the power spectrum from PBHs abundance bounds. Some possibilities of nonstandard Big Bang come from theories like supersymmetry \citep{PhysRevD261231, Dine:1995uk, Moroi:1999zb,Lyth:1995hj}, string theory  \citep{Cicoli:2023opf, Asaka:1999xd, Dutta:2016htz}, and modified gravity \citep{TDamour_1992, Cline:1999ts, Csaki:1999jh}. Previous studies have investigated particular expansion histories, addressing the formation of PBHs and their implications for the power spectrum \citep{Josan:2009qn, Cole:2017gle, Carr:2017edp, Bhattacharya:2021wnk, Bhattacharya:2019bvk}. In contrast, the present study takes a more general perspective on the conditions to form PBHs and how these evolve within non-standard scenarios, focusing on the modification of the constraints to the PBH abundance. \footnote{Throughout this paper, we consider that the PBHs formed are of the Schwarzschild kind, an assumption justified below.}. 

In the most general case, one could contemplate a sequence of non-standard stages with different equations of state, such as an early MD epoch, followed by an SD period, or indeed the QCD transition \citep{Bhattacharya:2023ztw}. However, in this work, we will focus only on simple scenarios where only one of these periods takes place at the end of inflation. 

The outline of this article is as follows. In section \ref{Sec2}, we review different scenarios that yield a nonstandard expansion of the universe. In section \ref{sec3}, we provide a summary of PBH formation during an RD epoch as well as different constraints on their abundance in the standard cosmology. In section \ref{sec4}, we focus on extending the existing constraints on PBHs for the particular cases of an early epoch of matter domination, matter domination triggered by a scalar field, and that of stiff fluid. For this purpose, we wrote the \href{https://github.com/TadeoDGAguilar/PBHBeta}{\faGithubSquare}~\texttt{PBHBeta} software, freely available at  \citep{PBHBeta_documentation, PBHBeta_repository}, which adapts the constraints on the abundance of PBHs for non-standard scenarios of evolution in the early universe. Section \ref{sec5} is devoted to conclusions.
 
\section{Nonstandard cosmologies}\label{Sec2}

{In the standard Big Bang model, it is typically assumed that after inflation, the universe quickly becomes RD, followed by an MD era from the equivalence redshift onwards. However, in this work, we are interested in non-standard evolution scenarios and in this section, we briefly review the motivation for the consideration of such epochs (for more detailed reviews, see \citep{2021OJAp....4E...1A, 2022arXiv221105767E})\footnote{Among the variety of scenarios that would allow deviations from the standard one, there could be extra inflation periods; e.g. the thermal inflation scenario \citep{PhysRevLett.75.201, PhysRevD.53.1784} or the ultra-slow roll inflation scenario \citep{Tsamis:2003px, Kinney:2005vj}. In these models, any PBHs formed during or before a period of inflation are expected to be diluted, resulting in a negligible abundance. Such cases are thus not considered here.}.}

\subsection{A fast-oscillating inflaton} 

At the end of inflation, it is expected that the inflaton field rolls to the minimum of its potential and stays oscillating around this minimum while it transfers energy to other (standard model) particles. This is the so-called reheating period \citep{PhysRevLett.73.3195,PhysRevD.56.3258,Allahverdi:2010xz,2019arXiv190704402L,RevModPhys.78.537,Amin2014eta}, which typically is considered to occur instantly, although the only restriction is that it must end before BBN. This gives way to considering an extended reheating process after inflation. In this scenario, the inflaton stays oscillating around its minimum during an extended period. 

During the fast oscillating regime of the inflaton, the minimum of the inflaton potential can be approximated by a power-law, $V(\varphi)\propto \varphi^{2n}$, and the (time-average) equation of state, $w$ will be determined by the power $n$ \citep{PhysRevD.28.1243}
\begin{equation}
w = \frac{n-1}{n+1}\,.
\end{equation}
With this expression, one can identify three scenarios: $n = 1,\,2$ and $n\neq 1,\,2$. For the case $n=1$, $w = 0$ which means that during reheating, the background universe effectively behaves as an MD universe. In this case, fluctuations yield a primordial structure formation process either through the formation of oscillons \citep{MustafaA.Amin_2010, PhysRevLett.108.241302, PhysRevD.83.096010, 2010arXiv1006.3075A} or a CDM-like structure formation \citep{Niemeyer:2019gab, PhysRevD.103.063525, Eggemeier:2021smj, Padilla:2021zgm, Hidalgo:2022yed,Padilla:2023lbv,Padilla:2024iyr}, In this process, PBHs may form as we shall see below in more detail.
The universe behaves effectively as an RD ($w = 1/3$) for $n = 2$. When $n\neq 1,2$, the equation of state will approach $w =1/3$ after enough time thanks to energy transfer from the homogeneous condensate to relativistic waves of the inflaton field \citep{PhysRevD.97.023533, PhysRevLett.119.061301}. This implies that if the reheating era takes place in a $n\neq 1$ stage, PBH formation will be similar to the standard cosmology.

\subsection{Heavy particles and dark sectors}

This scenario arose initially in the context of supersymmetry and string theory (see, for example \citep{MOROI2000455, PhysRevD.26.1231, COUGHLAN198359, PhysRevD.50.6357}). In the early universe, a quasi-stable massive particle, $\psi$ could dominate before BBN. These particles may or may not have been in thermal equilibrium with the rest of the standard model particles. Since these particles are massive, they are expected to become non-relativistic very quickly and behave collectively as pressureless dust. Even if these particles did not initially dominate at formation time, if they are sufficiently long-lived, they could dominate since the particle energy density ($\rho_\psi\propto a^{-3}$) decreases slower than the background radiation ($\rho_{\rm rad}\propto a^{-4}$). 
As a result, we should expect an early MD period in the universe. This period would end once the heavy particle decays into relativistic degrees of freedom \citep{2021OJAp....4E...1A}.

\subsection{Moduli fields}

Moduli fields are scalar fields that emerge in string theory and M-theory compactifications \citep{Cicoli:2023opf}. In a cosmological context, moduli fields are typically assumed to be in fast oscillations around the minimum of their potential, which provokes an early MD era \citep{Kane:2015jia, PhysRevD.80.083529}. 

These moduli fields can decay late enough to affect the BBN, which would cause a problem for these models, known as the cosmological moduli problem. On the contrary, if the moduli fields are massive enough, they could decay prior to BBN, provoking a second reheating period. In this situation, the moduli fields could have some effects on the dark matter, baryogenesis, or the structure formation process \citep{Kane:2015jia}.  

\subsection{Early stiff-dominated universe}

More generically, the universe may have been dominated by a particle species, $\psi$ with an equation of state, $1/3< w_\psi <1 $. In this scenario, the universe would initially be dominated by $\psi$ but because its energy density decays faster than radiation ($\rho_\psi \propto a^{-3(1+w_\psi)}$), it would become subdominant once enough time has passed. In this case, $\psi$ does not need to decay, so it can be completely stable.

This stiff epoch has several interesting implications for the early universe, such as the enhancement of the primordial gravitational wave background generated during inflation \citep{Li:2016mmc}. Equivalently, one would expect an increase in the density of PBHs during this early epoch. This will be studied in more detail below.

\section{Primordial Black Holes in Standard Cosmology}\label{sec3}

In this section, we briefly explain the properties of PBHs, their formation mechanism, and their observational constraints. We shall assume the standard Big Bang scenario. 

\subsection{Mass and fraction of PBHs}\label{massandfraction}

The standard scenario to account for PBHs is the following: if a perturbation at the time of reentry to the horizon exceeds a threshold value, $\delta_c$ it will collapse to form a PBH. This threshold value strongly depends on the morphology of the initial perturbation that will collapse to form the PBH \citep{Musco:2018rwt}. In the RD standard cosmology, the threshold value is \citep{Niemeyer:1997mt,Musco:2008hv,Harada:2013epa,Musco:2018rwt,PhysRevD.101.044022,PhysRevD.103.063538}
\begin{equation}
\delta_c^{(\rm RD)} = 0.41-0.66\,.
\end{equation}
In this manuscript, we will use $\delta_c^{\rm (RD)} = 0.41$ since this is the most commonly used value in the literature.

The mass of a PBH is equal to a fraction of the total mass of the horizon at the time of horizon crossing, 
$M_{\rm PBH} = \gamma\,M_{H}|_{k = aH}$. This depends on the background equation of state, from the number of $e-$folds elapsed to horizon reentry as a function of wave number $k$, and to the functional dependence of the mass on the wavenumber. For example, $\gamma$ is a constant that encrypts the efficiency of the collapse and for an RD era is $\gamma^{(\rm RD)} = (1/3)^{3/2}$ \citep{PhysRevD.81.104019}. We can express the mass, $M_{\rm PBH}$ as \citep{Hidalgo:2022yed}
\begin{equation}\label{mass}
\frac{M_{\rm PBH}(k)}{7.1\times 10^{-2}~\rm{g}} =  \gamma^{(\rm RD)}\frac{1.8\times 10^{15}~\rm{GeV}}{H_{\rm HC}(k)}\,,
\end{equation}
where subfix $_{\rm HC}$ is used to refer to quantities evaluated at the time of horizon crossing.

In standard cosmology, PBHs are assumed to form almost instantly after the scale of perturbation, $k$ reenters the horizon. Assuming adiabatic cosmic expansion, the fraction $\beta(M_{\rm PBH})\equiv \rho_{\rm PBH}(t_i)/\rho(t_i)$ of the mass of the universe collapsing into PBHs at their formation time, $t_i$ is given by \citep{Carr:2020gox} 
\begin{equation}\label{eq_betap}
\beta \simeq 7.06\times 10^{-18}\Omega^0_{\rm PBH}\left(\frac{M_{\rm PBH}}{10^{15}~\rm{g}}\right)^{1/2}\,.
\end{equation}
In the above expression 
\begin{equation}
\beta^{'}\equiv (\gamma^{(\rm RD)})^{1/2}\left(\frac{g_{*i}}{106.75}\right)^{-1/4}\left(\frac{h}{0.67}\right)^{-2}\beta\,,    \label{beta:betaprime}
\end{equation}
where $g_{*i}$ is the number of relativistic degrees of freedom at the formation time, $h$ is the reduced Hubble parameter, 
and $\Omega^{0}_{\rm PBH}\equiv \rho_{\rm PBH}(t_0)/\rho_{\rm crit}(t_0)$ is the current density parameter of the PBHs. Eqs.~\eqref{eq_betap} and \eqref{beta:betaprime} are, of course, dependent on the dominating matter, as we shall show in the next section.

\begin{figure}{H}
\centering
\includegraphics[width=3.5in]{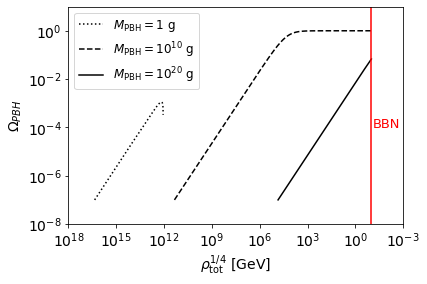}
\caption{\footnotesize{Abundance of PBHs as a function of the energy density and for the particular initial condition $\beta(M_{\rm PBH}) = 10^{-7}$. In each scenario, we evolved $\Omega_{\rm PBH}$ from the formation energy scale up to the BBN energy scale at $\rho_{\rm BBN}^{1/4} = 100~\rm{MeV}$.}}
\label{fig:evrd}
\end{figure}

Once PBHs form, they lose mass via Hawking radiation \citep{Hawking:1974rv} and their mass as a function of the cosmic time, $t$ changes as \citep{Dalianis:2021dbs}
\begin{equation}\label{teva}
  M_{\rm PBH}(t) = M_{\rm PBH}(t_{i})\left(1-\frac{t-t_i}{\Delta t_{\rm eva}}\right)^{1/3}\,,   
\end{equation}
where $\Delta t_{\rm eva}$ is the time at which a PBH evaporates completely
\begin{equation}
\Delta t_{\rm eva} \equiv t_{\rm eva}-t_i=t_{\rm Pl}\left(\frac{M_{\rm PBH}(t_{i})}{M_{\rm Pl}}\right)^{3}\,,
\end{equation}
and $t_{\rm Pl}$ ($M_{\rm Pl}$) is the Planck time (mass). Therefore, only PBHs with masses $M_{\rm PBH} \gtrsim 10^{15}$ g should prevail.
To calculate how this mass loss affects the abundance of PBHs, we can proceed as follows. If we define $\bar\beta$ as the mass fraction in the absence of Hawking radiation, we can express $\Omega_{\rm PBH}(M;t)$ as  \citep{Martin:2019nuw}
\begin{equation}\label{omx}
\Omega_{\rm PBH}(M_{\rm PBH};t) = \bar\beta(M_{\rm PBH};t)\left(1-\frac{t-t_i}{\Delta t_{\rm eva}}\right)^{1/3}\,.
\end{equation}
Following the description in \citep{Martin:2019nuw}, we decompose $\bar\beta(M_{\rm PBH};\,t)$ as $\bar\beta(M_{\rm PBH};\,t) = b(t)\bar\beta(M_{\rm PBH},\,t_{i})$, where $b(t)$ solves the differential equation 
\begin{equation}\label{b}
\dot b(t)+\left(\frac{\dot \rho_{\rm tot}}{\rho_{\rm tot}}+3H\right)b(t) = 0\,.
\end{equation}
Here $\rho_{\rm tot}$ is the total energy density of the background universe. In this work, we evolve this differential equation from the PBH formation time to the BBN epoch, which takes place at $\rho_{\rm BBN}^{1/4} = 100~\rm{MeV}$ (a fiducial value). In standard cosmology, we have $\rho_{\rm tot} \approx \rho_{\rm rad}+\rho_{\rm PBH}$ and $\rho_{\rm rad}(t_i)\gg \rho_{\rm PBH}(t_i)$. Thus we rewrite Eq.~\eqref{b} as
\begin{equation}\label{eqdb}
\frac{db}{d\ln \rho_{\rm tot}}+ b\,\frac{\Omega_{\rm{PBH}}-1}{\Omega_{\rm PBH}-4} = 0\,.
\end{equation}
To calculate the density parameter, Eq.~\eqref{omx} we need to compute $t-t_i$. This time is computed from the differential equation
\begin{equation}\label{eqdt}
\frac{d(t-t_i)}{d\ln \rho_{\rm tot}} = \frac{\sqrt{3}m_{\rm Pl}}{(\Omega_{\rm PBH}-4)\sqrt{\rho_{\rm tot}}}.
\end{equation}
Eqs.\,\eqref{omx}, \eqref{eqdb} and \eqref{eqdt} with appropriate initial conditions such as
\begin{equation}\label{inicon}
b(t_i) = 1\,, \ \ \ \  \bar \beta(M_{\rm PBH};\,t_i) = \beta(M_{\rm PBH})\,, 
\end{equation}
are the system of differential equations that give us the abundance of PBHs at a particular time, $t$. In~\autoref{fig:evrd} we show the evolution of $\Omega_{\rm PBH}$ as a function of the total energy density, $\rho_{\rm tot}$ for three different mass of PBHs, $M_{\rm PBH} = (1,\,10^{10}$,\,$10^{20})\,\rm{g}$. 
Note that in the case in which the density of PBHs becomes unimportant during the early universe ($\Omega_{\rm PBH}\ll 1$) from Eq.~\eqref{eqdb} we obtain $b\propto \rho_{\rm tot}^{-1/4}\propto a$. On the contrary, in the case PBHs dominate the early universe ($\Omega_{\rm PBH}\simeq 1$), we obtain $b = \mathrm{const}$. Finally, when $t\simeq \Delta t_{\rm eva}$ an abrupt drop in the density of PBHs is obtained. This system is valid only in the standard Big Bang. Therefore, a different evolution at early times will impact these results.\\ 

\subsection{Constraints on PBHs abundance}\label{constraints}

\begin{figure}[H]
\centering
\includegraphics[width=3.5in]{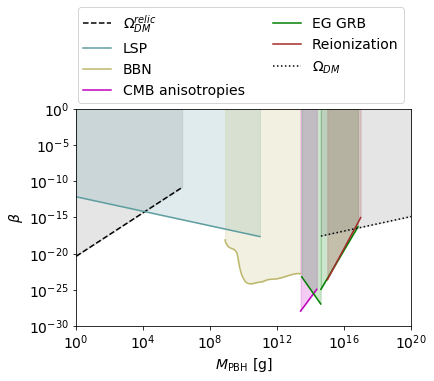}
\caption{\footnotesize{Constraints on the abundance of monochromatic PBHs as a function of mass in the standard Big Bang scenario. Observations exclude the shaded coloured regions.}}
\label{fig:beta_rad}
\end{figure}

In this section, we compile the most restrictive constraints on the abundance of PBHs with the assumption of the standard evolution of the universe. The main results are shown in\,\autoref{fig:beta_rad}. 

\textit{Dark matter.} PBHs with a mass larger than $10^{15}$\,g have a lifetime larger than the age of the universe and can be candidates for dark matter with maximum abundance $\Omega_{\rm CDM} = 0.264\pm 0.006$ \citep{Planck:2018jri}. From Eq.~\eqref{eq_betap}, this condition is met provided that (dotted black line in\,\autoref{fig:beta_rad})
\begin{equation}
\beta^{'}(M_{\rm PBH}) \lesssim 1.86\times 10^{-18}\left(\frac{M_{\rm PBH}}{10^{15}~\rm{g}}\right)^{1/2}\,.
\end{equation}
    
\textit{Big Bang Nucleosynthesis.} Evaporated PBHs can affect the abundances of light elements, specifically hydrogen, deuterium, and helium \citep{Keith:2020jww, PhysRevD.81.104019, Carr:2020gox}. PBHs in the mass range $6\times 10^8~\rm{g} \lesssim M_{\rm PBH}\lesssim 2\times 10^{13}~\rm{g}$ will evaporate during or after BBN, leading to an injection of particles that can impact the predictions of BBN. The constraints necessary for this injection not to alter BBN predictions can be found in \citep{Keith:2020jww, PhysRevD.81.104019, Carr:2020gox} and are shown in \autoref{fig:beta_rad} with a lime-green line. 

\textit{CMB anisotropies.} A constraint of PBHs in the mass range $2.5\times 10^{13}~\rm{g}\lesssim M_{\rm PBH}\lesssim 2.4\times 10^{14}~\rm{g}$ is associated with CMB physics. It comes from the damping of small-scale CMB anisotropies \citep{PhysRevD.81.104019}. 
\begin{equation}
\beta^{'}<3\times 10^{-30}\left(\frac{M_{\rm PBH}}{10^{13}~\rm g}\right)^{3.1}\,.
\end{equation}
This constriction is also shown in\,\autoref{fig:beta_rad} with a violet line. 

\textit{Extragalactic $\gamma$-ray background.} PBHs in the range of masses $3\times 10^{13}~\rm{g}\leq M_{\rm PBH}\leq 5.1\times 10^{14}~\rm{g}$, could contribute to the diffuse x-ray and $\gamma$-ray background \citep{PhysRevD.81.104019} and approximately fulfill the condition
\begin{equation}
\beta^{'}< 5\times 10^{-28}\left(\frac{M_{\rm PBH}}{5.1\times 10^{14}~\rm{g}}\right)^{-3.3}\,.
\label{eq:g_ray1}
\end{equation}
Similarly, PBHs with masses $5.1\times 10^{14}~\rm{g}\leq M_{\rm PBH}\leq 7\times 10^{16}~\rm{g}$, which are about to evaporate, fulfill
\begin{equation}
\beta^{'}< 5\times 10^{-26}\left(\frac{M_{\rm PBH}}{5.1\times 10^{14}~\rm{g}}\right)^{3.9}\,.
\label{eq:g_ray2}
\end{equation}
We plotted these limits with green lines in\,\autoref{fig:beta_rad}.

{In this work we do not discriminate between the first and second emission of particles in the Hawking radiation process, since the constraints imposed on the gamma-ray background consider the impact of the total radiation on the evaporation of PBHs. In fact, all of the constraints presented in \autoref{fig:beta_rad} take into account the addition of both emissions (see e.g. \citep{PhysRevD.41.3052, MacGibbon:1991vc}). Moreover, the evaporation rate is rarely affected by the spin of the emitting black hole \citep{Starobinsky:1973aij}, and thus we assume Schwarzschild black holes throughout.}

\textit{Reionization.} Ref.\,\citep{Clark:2016nst} imposed constraints on PBHs in the range of masses $10^{15}~\rm{g}<M_{\rm PBH}< 10^{17}~\rm{g}$. Hawking radiation from PBHs with a lifetime greater than the age of the universe leaves an imprint on the CMB through modifications of the ionization history and the damping of CMB anisotropies. This study resulted in the constraint 
\begin{equation}
\beta^{'}<2.4\times 10^{-26}\left(\frac{M_{\rm PBH}}{M_*}\right)^{4.3}\,.
\end{equation}
We show this limit with a brown line in~\autoref{fig:beta_rad}.

\textit{Lightest Supersymmetric Particles.} The evaporation process of PBHs can produce any other particles predicted in theories beyond the standard model of particle physics. In particular, the evaporation of PBHs may produce the lightest supersymmetric particles (LSP), predicted in supersymmetry and supergravity models, which are stable and may contribute to the totality of the dark matter in the universe. If these LSPs are produced, the condition for them not to exceed the CDM matter density imposes the bound \citep{LEMOINE2000333}
\begin{equation}
\beta^{'}(M_{\rm PBH})\lesssim 10^{-18}\left(\frac{M_{\rm PBH}}{10^{11}~\rm{g}}\right)^{-1/2}\left(\frac{m_{\rm LSP}}{100~\rm{GeV}}\right)^{-1}\,.
\end{equation}
The above expression is valid for PBHs with masses $M_{\rm PBH}<10^{11}(m_{\rm LSP}/100~\rm{GeV})^{-1}~\rm{g}$. We plotted this limit in~\autoref{fig:beta_rad} with a cyan line and for the particular example $m_{\rm LSP} = 100~\rm{GeV}$.

\textit{Planck Mass Relics.} If, after evaporation, PBHs leave stable Planck-mass relics, these might contribute to dark matter \citep{MacGibbon:1987my, 2007hep.th....3070A, Carr:2020gox}. If we assume relics have a mass $\kappa M_{\rm Pl}$, the constriction of the abundance of PBHs with masses $M_{\rm PBH}<10^{11}\kappa^{2/5}M_{\rm Pl}$ is
\begin{equation}
\beta^{'}< 8\times 10^{-28}\kappa^{-1}\left(\frac{M_{\rm PBH}}{10^9~\rm{g}}\right)^{3/2}\,,
\label{betaprime:evap}
\end{equation}
otherwise, PBHs would have a larger abundance than the totality of the dark matter in the universe. The upper mass limit arises because a population of PBHs with masses larger than this would dominate the total energy of the universe before their evaporation 
 even when meeting the bound of Eq.~\eqref{betaprime:evap}. This limit for $\kappa = 1$ is plotted with a dashed black line in~\autoref{fig:beta_rad}.

\subsection{Constraints on the power spectrum imposed by PBHs} 

According to the Press-Schechter formalism \cite{Press:1973iz}, the fraction of the energy density of the universe contained in a region with size $R$, which is overdense enough to form PBHs is given by \cite{Carr:1975qj} 
\begin{eqnarray}\label{PS2}
\beta(M_{\bh}) &=&\gamma \int_{\delta_c}^{\infty} \frac{d\delta}{\sqrt{2\pi\,\sigma^2_\delta(R)}} \, \exp\left(-\frac{\delta^2}{2\,\sigma^{2}_\delta(R)}\right)\nonumber\\
&=& \frac{\gamma}{2}\,\erfc\left(\frac{\delta_c}{\sqrt{2\sigma^{2}}}\right)\,,
\end{eqnarray}
where $\erfc(x) = 1 - \erfc(x)$ is the complementary error function and the variance of $\delta$ is given by
\begin{eqnarray}
\sigma^2(R) &=&\frac{4(1+\omega)^2}{(5+3\omega)^2} \int_0^{\infty}(kR)^2W^2(kR)\,\mathcal{P}_\zeta(k)\,\frac{dk}{k},\nonumber\\ 
&\simeq& \frac{4(1+\omega)^2}{(5+3\omega)^2}\mathcal{P}_\zeta(k).\label{sigma_square}
\end{eqnarray}
Here $W(k,\,R)$ is the Fourier transform of the window function used to smooth the density contrast on a comoving scale $R$.

Note that 
any constraint on $\beta(M_{\rm PBH})$ thus imposes a constraint on $\sigma$, and equivalently on power spectrum, $\mathcal{P}_\zeta (k)$. In such correspondence, the dependence on the background dominating matter lies both on the threshold amplitude, $\delta_c$ for PBH formation, and on the relation between the variance and the power spectrum.

\begin{figure}[H]
\centering    
\includegraphics[width=3.4in]{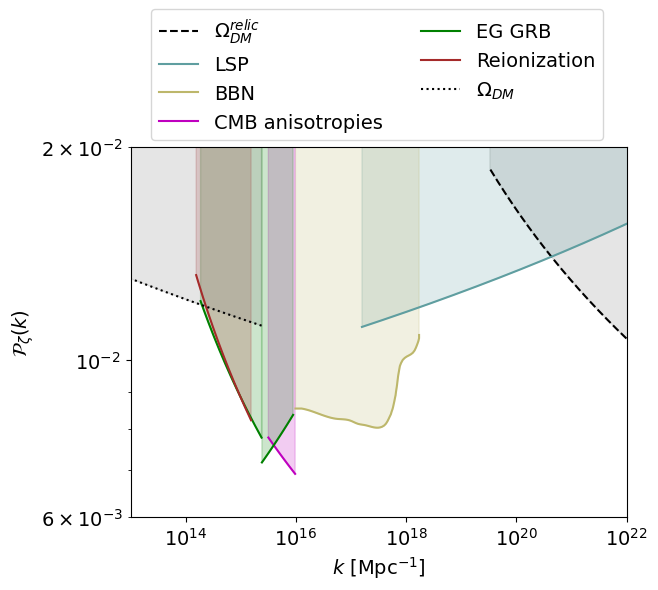}
\caption{\footnotesize{Constraints imposed on power spectrum, $\mathcal{P}_\zeta$ by PBHs in the standard Big Bang scenario.}}
\label{fig:my_label}
\end{figure}

In\,\autoref{fig:my_label} we plot different constraints reviewed in the subsection \ref{constraints}, but now applied to $\mathcal{P}_\zeta(k)$ and for the particular case in which $\delta_{c}^{\rm (RD)}=0.41$ \citep{Harada:2013epa, Niemeyer:1999ak, Hawke:2002rf, Musco:2008hv, Bhattacharya:2023ztw}.

\section{Primordial Black Holes in non-Standard Cosmology}\label{sec4}
In this section, we study PBHs formation and their abundance in scenarios that deviate from the standard Big Bang.

\subsection{Early matter dominated scenario}\label{evolutionmd}
As a first case in modifying the standard scenario, we shall consider an early MD phase before BBN\footnote{Even if the radiation does not have an initial negligible value, since the energy density of radiation dilutes faster than that of matter, it will quickly become negligible.}. Here, we consider a particular MD scenario in which the amount of radiation is a consequence of the decay of the hypothetical particle, $\psi$. This particle could be, as described in Sec.~\ref{Sec2}, either heavy, a particle associated with moduli fields, or the inflaton field itself. 
 We also assume that this MD era follows immediately after inflation and PBHs form during this era. 
 
 A few differences should be considered between this case and the standard instant reheating scenario (as discussed e.g. in \citep{Bhattacharya:2023ztw, Bhattacharya:2021wnk}). First, to determine the mass of a black hole from an inhomogeneity of size $2\pi/k$, we need to replace $\gamma^{(\rm RD)}$ in Eq.\,\eqref{mass} with $\gamma^{\rm (MD)}$, and determine the Hubble factor, $H_{\rm HC}(k)$ according to this non-standard cosmology. The particular value of $\gamma^{\rm (MD)}$ is not well known and we thus adopt $\gamma^{\rm (MD)} = 1$ \citep{Martin:2019nuw}. 

After PBH formation, the evolution of the density parameter, $\Omega_{\rm PBH}$ (and its relation to $\beta$), is modified with respect to the standard cosmology. The function $b(t)$ is determined through a system of differential equations equivalent to that in Eqs.~\eqref{omx}, \eqref{eqdb} and \eqref{eqdt}, with the new fluid $\Omega_{\psi}$ that dominates the universe; i.e.
\begin{eqnarray}\label{eq:db}
\frac{db}{d\ln \rho_{\rm tot}}+b\,\frac{\Omega_{\rm{PBH}}+\Omega_{\psi}-1}{\Omega_{\rm PBH}+\Omega_{\psi}-4} = 0\,.\\
\dfrac{d\,\Omega_{\psi}}{d\ln \rho_{\rm tot}}+ \Omega_{\psi}\,\frac{\Omega_{\rm{PBH}}+\Omega_{\psi}-1}{\Omega_{\rm PBH}+\Omega_{\psi}-4} = 0\,.\\
\dfrac{d(t-t_i)}{d\ln \rho_{\rm tot}} = \frac{\sqrt{3}m_{\rm Pl}}{(\Omega_{\rm PBH}+\Omega_{\psi}-4)\sqrt{\rho_{\rm tot}}}\,.\label{eq:dt}
\end{eqnarray}
These differential equations should be evolved with adequate initial conditions up to the energy scale at which the particle, $\psi$ decays; i.e. $\rho_{\rm tot}^{\rm (dec)} > 
 \rho_{\rm BBN}$. We choose to set initial conditions similar to the standard cosmology (Eq.\,\eqref{inicon}), and in addition, we have
\begin{equation}
\Omega_{\psi}(t_i)=1-\beta(M_{\rm PBH})\,.
\end{equation}
The above condition implies a negligible contribution from radiation at the initial time, $t_i$. 

\begin{figure}[H]
\centering
\includegraphics[width=3.5in]{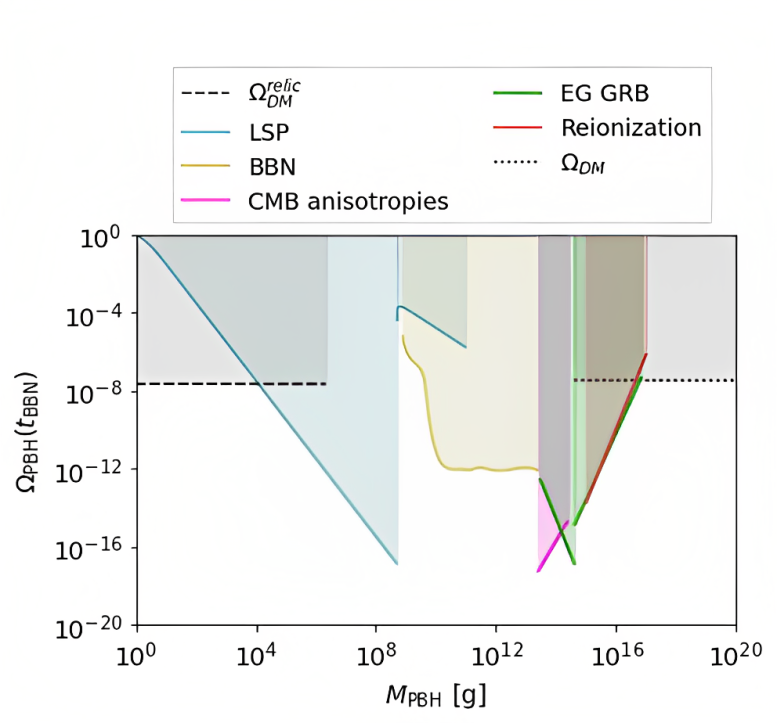}
\caption{\footnotesize{Constraints on the abundance of PBHs at BBN. At this BBN epoch, PBHs with masses $M_{\rm PBH}\lesssim 10^9~\rm{g}$ have already evaporated, and their constraints are imposed following the evolution of the Planck mass remanents of the evaporation.}}
\label{fig:omrad}
\end{figure}

Once the energy scale, $\rho_{\rm tot}^{\rm (dec)}$ is reached, the energy associated with the particle, $\psi$ is transferred to the radiation background ($\Omega_\psi(t_{\rm dec})\ \rightarrow \Omega_{\rm rad}(t_{\rm dec})$). Subsequently, the evolution from $\rho_{\rm tot }^{\rm (dec)}$ up to BBN is dictated by the system of Eqs.\,\eqref{omx}, \eqref{eqdb} and \eqref{eqdt} with the initial conditions imposed at $\rho_{\rm tot}^{\rm (dec)}$.

\begin{figure}[t]
\centering
\includegraphics[width=3.3in]{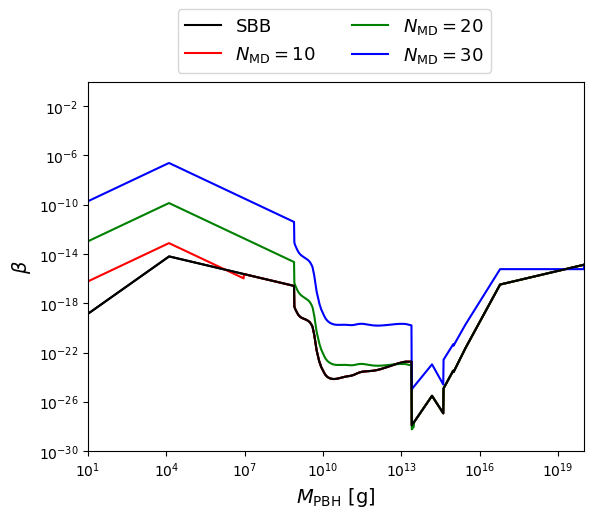}
\caption{\footnotesize{Constraints on the abundance of PBHs as a function of their mass. $N_{\rm MD}$ is the total number of $e$-folds that the MD epoch lasted. }}
\label{fig:const_md}
\end{figure}

Now, we analyze how this early MD epoch will modify the constraints on the abundance of PBH at the time of formation. The set of constraints reviewed in Sec.~\ref{constraints} (summarized in~\autoref{fig:beta_rad}) can be evolved up to the BBN scale, $\rho_{\rm BBN}$ by solving Eqs.\,\eqref{omx}, \eqref{eqdb}, and \eqref{eqdt}. These extrapolated constraints are shown in~\autoref{fig:omrad}. Note that, in particular, the constraint from the PBH component (or its relics) constituting the totality of dark matter imposes a constant maximum value (a horizontal line in~\ref{fig:omrad}) for each monochromatic mass population (as expected). On the other hand, if PBHs evaporate during or after BBN, we expect that the constraints given at the beginning of BBN should be the same in any scenario of previous evolution (radiation or any other matter domination). This is since the evolution of PBHs from BBN up to the present time (or up to the time they evaporate) is the same in any of the scenarios because we contemplate deviations from the standard cosmology only at energy scales larger than BBN. {A similar argument applies to the remnants of PBHs.} We thus extend the PBHs constraints to the early MD scenario in a {shooting method} as follows:
\begin{enumerate}
    \item We assume the constraints in\,\autoref{fig:omrad}, set at $t_{\rm BBN}$, are valid for any background evolution before this time.
    \item We consider different initial conditions on the abundance of PBHs at their formation time and evolve them through the formalism given in Eqs.~\eqref{eq:db} through \eqref{eq:dt}, up to the energy scale, $\rho_{\rm tot}^{1/4} = 100~\rm{MeV}$.
    \item For each mass, we look for the initial condition on $\beta(M_{\rm PBH})$ which meets the smallest of values reported in~\autoref{fig:omrad}.~This thus represents the constraints on the abundance of PBHs at the time of their formation in our non-standard scenario.
\end{enumerate}
The resulting constraints on PBH abundances for different durations of the MD period are shown in~\autoref{fig:const_md}. It is clear that the early MD era relaxes the constraints for small masses and converge to the constraints of standard cosmology for large masses. The reason is that, contrary to PBH formation in an RD epoch where the density of PBHs grows proportional to the scale factor, in the early MD scenario, their density remains constant and only enhances when the RD epoch starts. Additionally, the slope of the constraints at small masses is altered since the background evolution is modified. Such effect is, of course, present in all other modifications of the expansion history as described throughout the paper\footnote{The meticulous reader will also note a mismatch at the end of the MD constraints and the (blue) RD line at which constraints should converge. This is simply due to the difference in the numerical value of $\gamma^{(\rm MD)}$ and its RD equivalent. A more detailed study of the collapse of configurations, with a smooth transition between different eras, would have to look at a continuous change in the value of this factor.}. 
\begin{figure}[h]
\centering
\includegraphics[width=3.2 in]{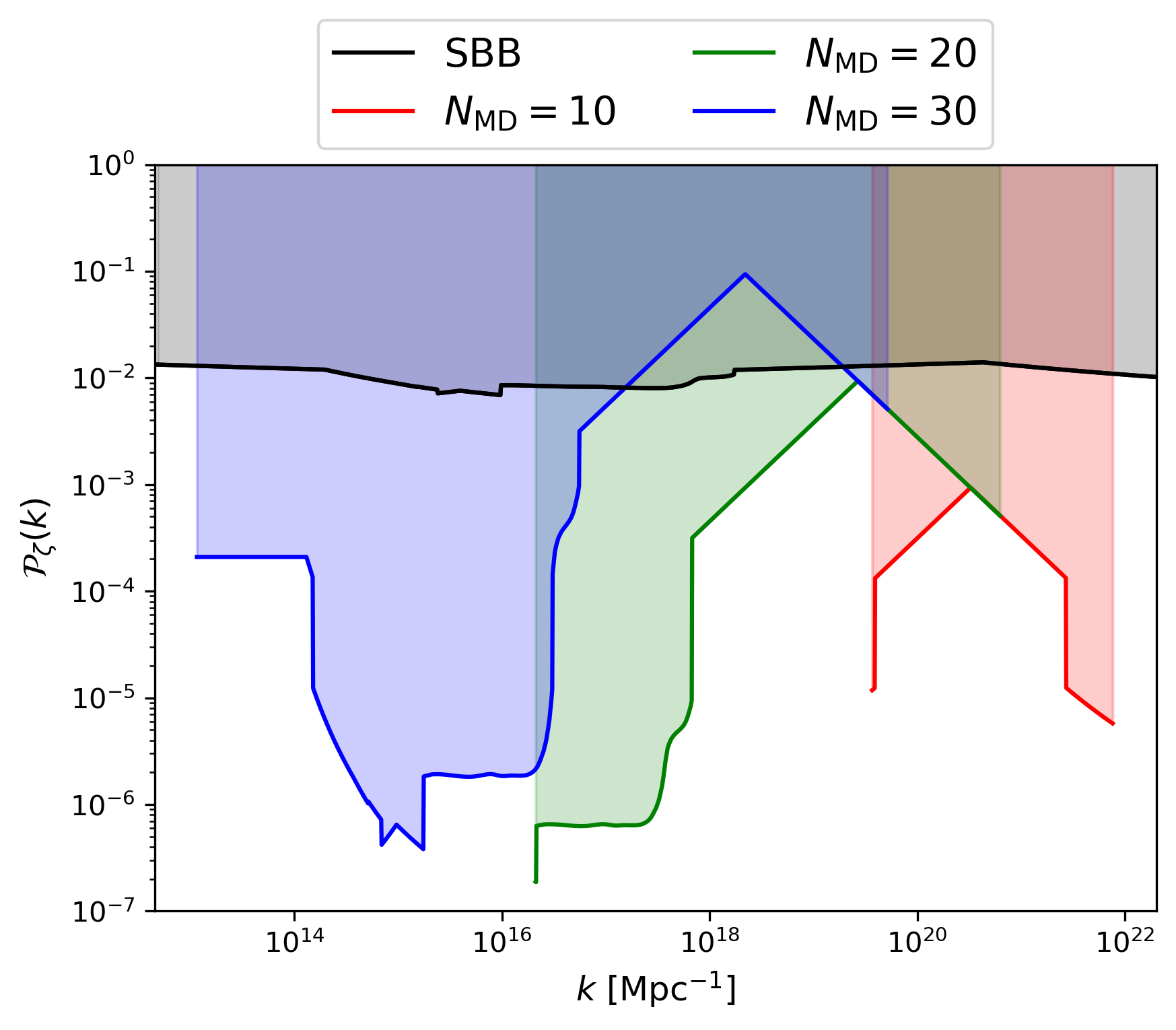}
\caption{\footnotesize{Constraints on $\mathcal{P}_\zeta(k)$ as a function of $k$ for different values of $N_{\rm MD}$. The jump in lines for $N_{\rm MD} = 10,\ 20$ coincides with the mass of the cosmological horizon at the time reheating ends.}}
\label{fig:S2_md}
\end{figure}
In~\autoref{fig:S2_md} we show the constraints on the power spectrum, $\mathcal{P}_\zeta(k)$, for a few values of the total number of $e-$folds of MD, $N_{\rm MD}$. To derive such constraints, instead of following the Press-Schechter formalism, we consider that the physical restrictions {like the effects of inhomogeneity and anisotropy  determine the relation between $\beta$ and $\sigma$ as follows \citep{Harada:2016mhb, Kokubu:2018fxy}
\begin{equation}\label{beta:sphere}
    \beta \simeq 0.2055\sigma^{13/2}, \quad \text{for} \quad  0.005\lesssim \sigma\lesssim 0.2\,.
\end{equation}
}
 Additionally, in the regime $\sigma < 0.005$, the effects of angular momentum of primordial fluctuations become important. The consequence is that the PBH probability is dictated by \citep{Harada:2017fjm}
\begin{equation}\label{beta:I}
\beta \simeq 1.9\times 10^{-7}f_q(q_c)\mathcal{I}^6\sigma^2 \exp(-0.15\mathcal{I}^{4/3}\sigma^{-2/3})\,,
\end{equation}
where $\mathcal{I}$ is a parameter of order $\mathcal{O}(1)$ and $f_{q}(q_{c})$ is the fraction of mass with a level of quadrupolar a sphericity, $q$ which is smaller than a threshold, $q_{c}$ (following Ref.\,\citep{Harada:2017fjm}, in our estimations we assume both $f_{q}(q_{c})$ and $ \mathcal{I}$ equal to one). It is clear that the formation of PBHs in an early MD scenario is much more likely than in an RD case, and thus, the bounds on the power spectrum are more restrictive than in the standard cosmology as shown in~\autoref{fig:S2_md}. {Additional restrictions can also be added to the collapse. Specifically, the inhomogeneities and anisotropies within a collapsing configuration \cite{Khlopov:1980mg, Harada:2016mhb, Kokubu:2018fxy}, or the dispersion velocity of particles \cite{Harada:2022xjp}, would modify the threshold amplitude of the collapse, as well as the abundance of PBHs. However, there are currently no general restrictions on $\delta_c$ or $\beta$ that may arise from such effects.} \\
{The careful reader may also wonder if a large abundance of PBHs would give room to the cloud-in-cloud problem \cite{Bardeen:1985tr}. It is worth noting that PBHs form at small scales and the cloud-in-cloud problem is relevant for large-scale structures like galaxies and halos. However, since PBHs are anti-correlated at short distances, they do not cluster. This is due to the fact that the formation of PBHs is an extremely rare event, thus  no such problem arises \cite{2024arXiv240200600A}. (see also \cite{Erfani:2021rmw}).}


\subsection{Scalar field dominated scenario}

Although at the level of cosmological background, a fast oscillating scalar field around the minimum of its potential behaves as a system of non-relativistic particles, i.e. (MD epoch), on small scales (of the order of the de Broglie wavelength associated with the scalar field), there is a different behavior than an MD universe. In particular, a quantum pressure effect is expected due to Heisenberg's uncertainty principle at small scales \citep{Schunck:2003kk}. This can help to stabilize structures and avoid the direct collapse of PBHs. It is well known that this quantum pressure generated by the scalar field (inflaton or a moduli field) would make it possible to form structures such as soliton-like structures due to the Bose-Einstein condensation \citep{Padilla:2021zgm, Hidalgo:2022yed,Padilla:2023lbv}. Furthermore, it was also shown that this quantum force can lead to the formation of similar CDM-like structures with a central soliton \citep{Dawoodbhoy:2021beb}. In this section, we consider the idea of including this quantum pressure effect, closely following \citep{Padilla:2021zgm, Hidalgo:2022yed,Padilla:2023lbv}, and we mention under what circumstances we could have the formation of PBHs.

After horizon crossing, it is expected that the structures relax through a violent relaxation mechanism and form CDM-like structures. 
If these configurations are massive enough, their size could be equal to or smaller than the Schwarzschild radius associated with the structure, so these configurations should collapse and form PBHs with a mass equal to Eq.\,\eqref{mass} (but with $\gamma^{\rm (MD)}$ instead of $\gamma^{\rm (RD)}$). The condition in terms of the density contrast sets the minimum amplitude for collapse as \citep{Padilla:2021zgm} 
\begin{equation}\label{delta_ih}
\delta_c^{(\rm CDM)} = 0.238\,.
\end{equation}

On the other hand, at the center of these CDM-like structures, the formation of a soliton profile with mass
\begin{equation}\label{m_sol}
    \frac{M_{\rm sol}(k)}{2.4\times 10^{-5}~\rm{g}} = \sqrt{\frac{\delta_{\rm HC}(k)}{1.39}}\frac{\rho_{11}^{1/6}(a_{\rm HC})}{\mu_5}\left(\frac{M_{\rm CDM}(k)}{7.1\times 10^{-2}~\rm{g}}\right)^{1/3}
\end{equation}
is expected, where $\rho_{11}(a_{\rm HC})\equiv 200\rho(a_{\rm HC})/(10^{11}~\rm{GeV})^4$, $\rho(a_{\rm HC})$ is the background density at the horizon crossing, $\mu_5 \equiv \mu/(10^{-5}m_{\rm Pl})$ and $\mu$ is an effective mass of the scalar field.
The number of $e$-folds necessary for the formation of such a soliton is given by
\begin{equation}
N_{\rm sol}(k) = N_{\rm NL}(k)+\frac{2}{3}\ln\left(1+\frac{\Delta t_{\rm cond}(k)}{t_{\rm NL}(k)}\right)\,,
\end{equation}
where
\begin{equation}\label{t_cond}
\frac{\Delta t_{\rm{cond}}(k)}{t_{\rm NL}(k)} = 8.159\times 10^{-18}\left(\mu_5^2M_{\rm CDM}(k)R_{\rm CDM}(k)\right)^{1/3}\,.
\end{equation}

\noindent In the above expression, $M_{\rm CDM}$ ($R_{\rm CDM}$) is the mass (radius) of the host CDM-like structure.
The threshold value for a soliton to collapse and form a PBH is \citep{Hidalgo:2022yed, Padilla:2021zgm}
\begin{equation}\label{delta_sol}
\delta_c^{(\rm sol)} = 0.019.
\end{equation}

\noindent After the formation of PBHs, their evolution will follow the description of the early MD era. Note that the time of formation of PBHs due to the gravitational collapse of soliton structures would be longer than the direct collapse scenario. Therefore, if the scalar field oscillation does not last long enough, this class of PBHs would not form\footnote{Ref.~\citep{DeLuca:2021pls} explored the possibility that structures like halos or solitons with a mass below their critical collapse threshold would grow in size through the process of accretion up to the critical mass required to collapse and form PBHs. Here, we do not consider such a possibility since it usually requires a longer reheating period beyond the already rather long reheating here considered.}.

Let us now look at the constraints on PBHs in this scenario. Since the time of formation of PBHs due to the gravitational collapse of CDM-like structures is the same as the ones that form in the direct collapse, the constraints on $\beta(M_{\rm PBH})$ should be the same as the ones shown in\,\autoref{fig:const_md}. On the other hand, in the case of PBH formation through the collapse of soliton-like structures, we expect from Eqs.\,\eqref{m_sol} and \eqref{delta_sol} that PBHs that can form more abundantly are the ones with a mass\footnote{Assuming that the amplitude of inhomogeneities forming PBHs take the critical value of Eq.~\eqref{delta_sol}.}
\begin{equation}
M_{\rm PBH} \simeq \frac{2.85\times 10^{-2}}{\mu_5}~\rm{g}\,.
\end{equation}
Therefore, PBHs from this scenario are very light and the relevant constraints are those of the Planck mass relics. 

\begin{figure}[H]
\centering
\includegraphics[width=3.4in]{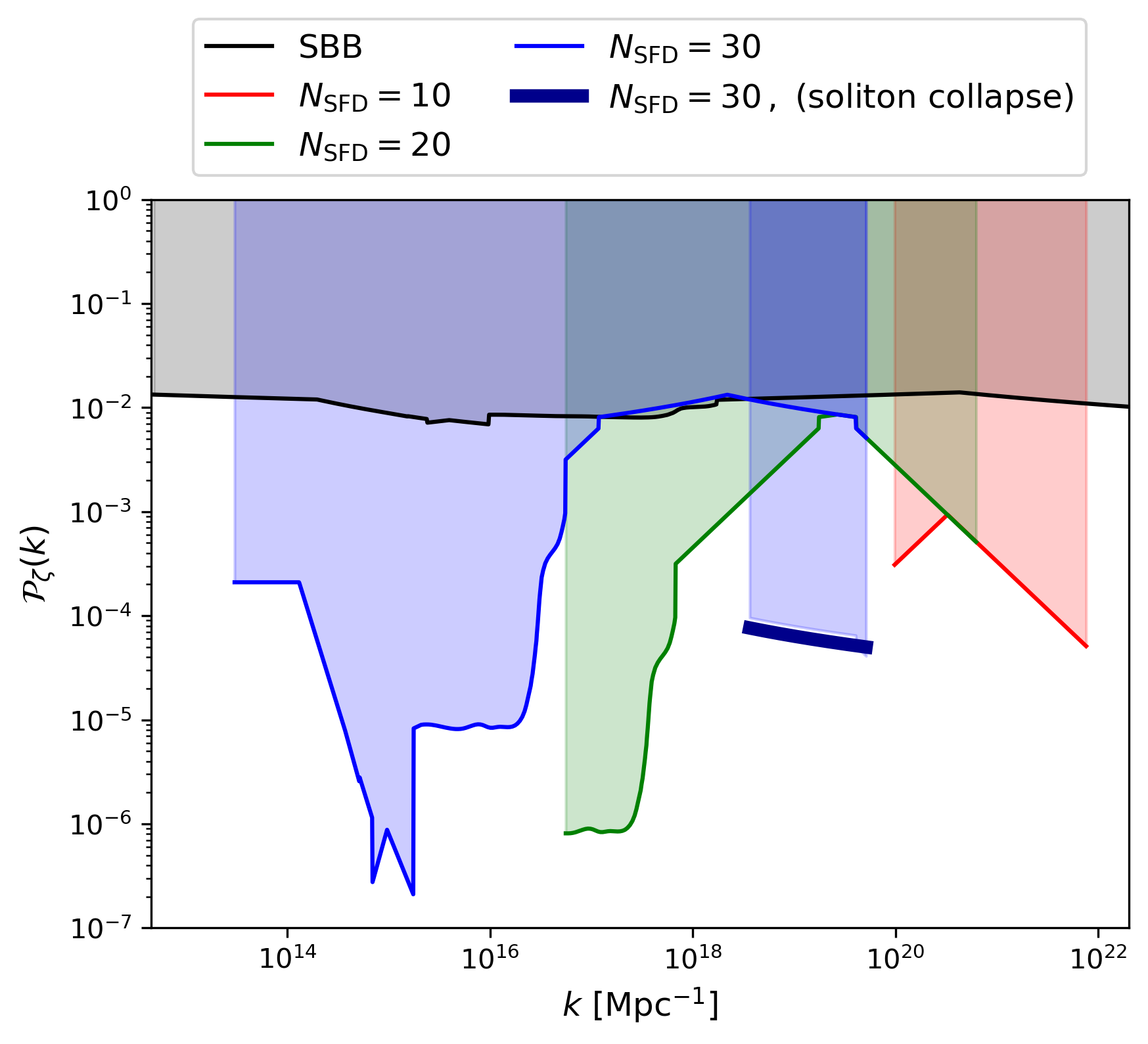}
\caption{\footnotesize{Different constraints on $\mathcal{P}_\zeta(k)$ as a function of $k$. The solid lines are the constraints that follow from the collapse of the CDM-like structures. The dashed line is the constraint that applies to the collapse of the soliton-like structures. We considered the region of parameters $\mu_5 \simeq 10^{-5}-1$.}}
\label{fig:psf}
\end{figure}

Although the constraints on $\beta$ are similar to those obtained by the early MD case, the correspondence with the power spectrum is significantly altered,  
since the PBH formation criterion is different.
In\,\autoref{fig:psf} we show the constraints considering this scenario of PBH formation. 
In this plot, we assume that the formation of a central soliton lasts approximately 30 $e$-folds (for more details see Refs.~\citep{Hidalgo:2022yed,Padilla:2024iyr}).

\subsection{Stiff fluid dominated scenario}

If the equation of state in the early universe becomes stiffer, the threshold value for PBH formation increases because, since pressure is higher, it is harder to collapse and form PBHs. In the limiting case of an equation of state, $w = 1$ we have \citep{Dalianis:2021dbs}
\begin{equation}\label{dc_sd}
\delta_c^{\rm (SD)} = 0.52-0.75\,,
\end{equation}
and we will take $\delta_c^{\rm (SD)} = 0.52$.

In an SD scenario, the mass of PBHs is given by Eq.\eqref{mass} by replacing $\gamma^{(\rm RD)}$ with $\gamma^{(\rm SD)}$. The particular value of $\gamma^{\rm (SD)}$ is not well known and in this paper we shall take $\gamma^{\rm (SD)}= 1$.
The evolution of the population of PBHs after their formation is described by the following equations
\begin{eqnarray}\label{eqs:sd}
\Omega_{\rm PBH}(M_{\rm PBH};t) = \bar\beta(M_{\rm PBH};t)\left(1-\frac{t-t_i}{\Delta t_{\rm eva}}\right)^{1/3}\,,\\
\frac{db}{d\ln \rho_{\rm tot}}+b\,\frac{\Omega_{\rm{PBH}}-2\Omega_{\psi}-1}{\Omega_{\rm PBH}-2\Omega_{\psi}-4} = 0\,,\\
\frac{d\Omega_{\psi}}{d\ln \rho_{\rm tot}}+ \Omega_{\psi}\,\frac{\Omega_{\rm{PBH}}-2\Omega_{\psi}+2}{\Omega_{\rm PBH}-2\Omega_{\psi}-4} = 0\,,\\
\frac{d(t-t_i)}{d\ln \rho_{\rm tot}} = \frac{\sqrt{3}m_{\rm Pl}}{(\Omega_{\rm PBH}-2\Omega_{\psi}-4)\sqrt{\rho_{\rm tot}}}\,.
\end{eqnarray}
We evolve this system of differential equations with the following initial conditions to obtain constraints on PBHs, 
\begin{eqnarray}
b(t_i) = 1\,,\ \ \ \ \ \bar \beta(M_{\rm PBH};t_i) = \beta(M_{\rm PBH})\,,\\ 
\Omega_{\psi}(t_i)=\frac{1-\beta(M_{\rm PBH})\left[1+\left(\frac{\rho_{\rm tot}(a_i)}{\rho_{\rm tot}(a_{\rm stiff})}\right)^{1/6}\right]}{1+\left(\frac{\rho_{\rm tot}(a_{\rm stiff})}{\rho_{\rm tot}(a_i)}\right)^{1/3}}\,,
\end{eqnarray}
where $\rho_{\rm tot}(a_i)$ and $\rho_{\rm tot}(a_{\rm stiff})$ are values of the density background at the formation and at the end of the stiff epoch\footnote{We consider the end of the stiff epoch when the radiation background equals the density of the stiff fluid.}, respectively.

\begin{figure}[H]
\centering
\includegraphics[width=3.3in]{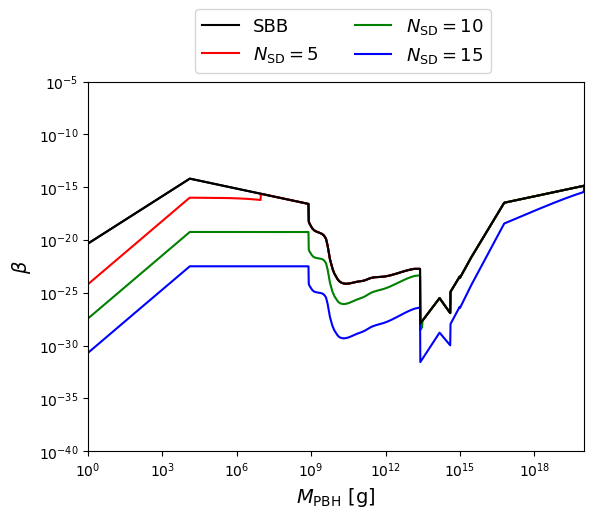}
\caption{\footnotesize{Abundance of PBHs as function of their mass, where $N_{\rm SD}$ is the total number of $e$-folds that the stiff era lasted.}}
\label{fig:bsd}
\end{figure}
\begin{figure}[H]
\centering\includegraphics[width=3.4in]{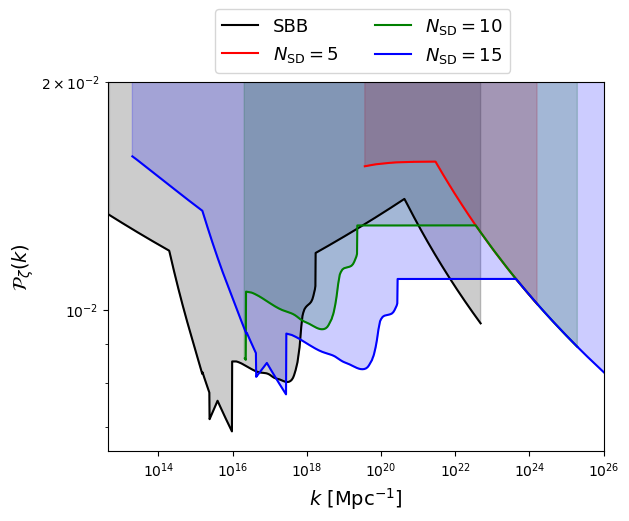}
\caption{\footnotesize{Constraints on $\mathcal{P}_\zeta(k)$ as a function of $k$ for the extended SD scenario.}}
\label{fig:psd}
\end{figure}

We show the constraints on the abundance of PBHs and the power spectrum in Figures\,\ref{fig:bsd} and \ref{fig:psd}, respectively. The early SD scenario leads to stronger constraints on the abundance of PBHs for the smaller masses. The reason is that, during the early SD scenario, the abundance of PBHs grows $\sim$$a^{3}$, which causes a much lower abundance in PBHs is needed to be able to comply with the constraints given in~\autoref{fig:omrad}.~This decrease in the constraints is despite the fact that the threshold value for the formation of PBHs is higher in the SD scenario than the standard Big Bang. Thus more restrictive values in the power spectrum are found for some values of the duration of the SD era, and only for the small masses at the end of the spectrum.

\section{Conclusions and discussions}\label{sec5}

In this work, we have studied the observational bounds to the abundance of Primordial Black Holes (PBHs), typically discussed in the context of a radiation-dominated universe following inflation immediately (the Standard Big Bang). Instead, we focused on a period of non-standard evolution in the early universe between inflation and the standard Big Bang. To account for the modifications of constraints on monochromatic populations of PBHs due to both the non-standard expansion and the specific criteria of PBH formation in each scenario, we developed the library \href{https://github.com/TadeoDGAguilar/PBHBeta}{\faGithubSquare}\,\texttt{PBHBeta}, freely available at \citep{PBHBeta_documentation, PBHBeta_repository}. 

As discussed, the modifications of constraints to the power spectrum are case-dependent. Our software considers up-to-date physical considerations, and the presented results are only examples of the general modifications that depart from the standard Big Bang yield.

Regarding the specific scenarios, we explored how the constraints on the power spectrum are modified in three different cases: 1) matter-dominated (MD) scenario, 2) scalar field-dominated ($\varphi$D) universe, common in reheating models, and 3) stiff fluid-dominated (SD) scenario. The results for these examples are shown in Figures~\ref{fig:const_md} through \ref{fig:psd}. Figures~\ref{fig:const_md} and \ref{fig:S2_md} display the constraints on the abundance of PBHs and the power spectrum in the early MD scenario, while Figures~\ref{fig:bsd} and \ref{fig:psd} do so for the case of an early SD scenario. The interest in the SFDM case arises because a quantum force is expected to prevent direct collapse and instead form soliton-like and primordial halo structures. Such structures can then collapse onto PBHs~\citep{Padilla:2021zgm}. This formation mechanism affects the constraints to the power spectrum (see~\autoref{fig:psf}). The changes to such constraints can be more stringent by up to two orders of magnitude in the relevant scales.

Note that since both the matter domination and scalar field domination cases exhibit the same background expansion, the constraints on the abundance of PBHs, $\beta(M)$ are the same in both scenarios. However, as discussed, \JCH{some of} the criteria for PBH formation differ, resulting in distinct bounds on power spectra, which is evident when comparing Figures \ref{fig:S2_md} and \ref{fig:psf}. Another important difference is the modification of constraints based on the duration of the non-standard expansion measured in $e$-foldings. The duration of non-standard expansion phases alters the correspondence between mass and wavenumber $k$, shifting the set of constraints to different $k-$values from the reference standard Big Bang case. 

Moreover, longer periods of MD weaken the constraints on the power spectrum. Conversely, in the SD case, where the PBH population redshifts as matter while the background evolves as the sixth power of the redshift, the abundance is enhanced. Nevertheless, the difficulty in forming PBHs in a stiff-dominated background means that the constraints on power spectrum are of the order of the standard Big Bang case throughout the PBH mass spectrum.

An interesting extension for this work will be to explore the implications of time-dependent equations of state for the background, which may represent more realistic scenarios. For example, realistic models of the QCD transition, e.g.~\citep{Cicoli:2023opf} and \citep{Burgess:2010bz}, suggest a time-dependent equation of state, which is well-motivated by models of string cosmology. It is important to assess whether  PBHs could impose constraints on these models. Additionally, one could explore the production of gravitational waves and their evolution within nonstandard cosmologies (see for example \citep{Domenech:2019quo, Domenech:2020kqm, Domenech:2021ztg}) or gravitational waves produced by the collision of PBHs, which persist after black hole evaporation. Such observables can be integrated into our set of constraints on ultra-light mass PBHs \citep{Papanikolaou:2020qtd, Papanikolaou:2022chm}. {On the other hand, the constraints on Power Spectra, derived here through the Press Schechter formula, may be computed by employing alternative PBH counting methodologies. This is the case of the Compaction Function \cite{Germani:2019zez}, a choice to be implemented in future versions of the \href{https://github.com/TadeoDGAguilar/PBHBeta}{\faGithubSquare}~\texttt{PBHBeta} package.} These and other extensions will be addressed in future work.\\


\section{Acknowledgements}

We are grateful to Nikita Blinov for providing the BBN constraints to the PBH abundance and the useful discussions. EE acknowledges the support of the ICTP, Italy for its hospitality and financial support from Dec.~2022 to Jun.~2023 when part of this work has been done. The authors acknowledge support from program UNAM-PAPIIT, grant IG102123 ``Laboratorio de Modelos y Datos (LAMOD) para proyectos de Investigación Científica: Censos Astrofísicos". TDG acknowledges financial support from CONAHCyT doctoral fellowship. LEP and JCH acknowledge sponsorship from CONAHCyT
Network Project No.~304001 ``Estudio
de campos escalares con aplicaciones en cosmolog\'ia y
astrof\'isica'', and through grant CB-2016-282569.  The work of
LEP is also supported by the DGAPA-UNAM postdoctoral grants program.

\bibliographystyle{ieeetr}

\begin{thebibliography}{100}

\bibitem{zel1967hypothesis}
Y.~B. Zel'dovich and I.~D. Novikov, ``The hypothesis of cores retarded during expansion and the hot cosmological model,'' {\em Soviet Astronomy}, vol.~10, p.~602, 1967.

\bibitem{Hawking:1974rv}
S.~W. Hawking, ``{Black hole explosions},'' {\em Nature}, vol.~248, pp.~30--31, 1974.

\bibitem{Page:1976wx}
D.~N. Page and S.~W. Hawking, ``{Gamma rays from primordial black holes},'' {\em Astrophys. J.}, vol.~206, pp.~1--7, 1976.

\bibitem{refId0}
{Lehoucq, R.}, {Cass\'e, M.}, {Casandjian, J.-M.}, and {Grenier, I.}, ``New constraints on the primordial black hole number density from galactic astronomy,'' {\em A\&A}, vol.~502, no.~1, pp.~37--43, 2009.

\bibitem{Cline_1997}
D.~B. Cline, D.~A. Sanders, and W.~Hong, ``Further evidence for some gamma-ray bursts consistent with primordial black hole evaporation,'' {\em The Astrophysical Journal}, vol.~486, p.~169, sep 1997.

\bibitem{PhysRevD.66.063505}
R.~Bean and J.~a. Magueijo, ``Could supermassive black holes be quintessential primordial black holes?,'' {\em Phys. Rev. D}, vol.~66, p.~063505, Sep 2002.

\bibitem{sasaki2016primordial}
M.~Sasaki, T.~Suyama, T.~Tanaka, and S.~Yokoyama, ``Primordial black hole scenario for the gravitational-wave event gw150914,'' {\em Phys. Rev. Lett.}, vol.~117, no.~6, p.~061101, 2016.

\bibitem{Harada:2013epa}
T.~Harada, C.-M. Yoo, and K.~Kohri, ``{Threshold of primordial black hole formation},'' {\em Phys. Rev. D}, vol.~88, no.~8, p.~084051, 2013.
\newblock [Erratum: Phys.Rev.D 89, 029903 (2014)].

\bibitem{PhysRevLett.73.3195}
L.~Kofman, A.~Linde, and A.~A. Starobinsky, ``Reheating after inflation,'' {\em Phys. Rev. Lett.}, vol.~73, pp.~3195--3198, Dec 1994.

\bibitem{PhysRevD.56.3258}
L.~Kofman, A.~Linde, and A.~A. Starobinsky, ``Towards the theory of reheating after inflation,'' {\em Phys. Rev. D}, vol.~56, pp.~3258--3295, Sep 1997.

\bibitem{Allahverdi:2010xz}
R.~Allahverdi, R.~Brandenberger, F.-Y. Cyr-Racine, and A.~Mazumdar, ``{Reheating in Inflationary Cosmology: Theory and Applications},'' {\em Ann. Rev. Nucl. Part. Sci.}, vol.~60, pp.~27--51, 2010.

\bibitem{2019arXiv190704402L}
K.~D. {Lozanov}, ``{Lectures on Reheating after Inflation},'' {\em arXiv e-prints}, p.~arXiv:1907.04402, July 2019.

\bibitem{RevModPhys.78.537}
B.~A. Bassett, S.~Tsujikawa, and D.~Wands, ``Inflation dynamics and reheating,'' {\em Rev. Mod. Phys.}, vol.~78, pp.~537--589, May 2006.

\bibitem{Amin2014eta}
M.~A. Amin, M.~P. Hertzberg, D.~I. Kaiser, and J.~Karouby, ``{Nonperturbative Dynamics Of Reheating After Inflation: A Review},'' {\em Int. J. Mod. Phys. D}, vol.~24, p.~1530003, 2014.

\bibitem{Nanopoulos:1979gx}
D.~V. Nanopoulos and S.~Weinberg, ``{Mechanisms for Cosmological Baryon Production},'' {\em Phys. Rev. D}, vol.~20, p.~2484, 1979.

\bibitem{Khlopov:1980mg}
M.~Y. Khlopov and A.~G. Polnarev, ``{PRIMORDIAL BLACK HOLES AS A COSMOLOGICAL TEST OF GRAND UNIFICATION},'' {\em Phys. Lett. B}, vol.~97, pp.~383--387, 1980.

\bibitem{Planck:2018jri}
Y.~Akrami {\em et~al.}, ``{Planck 2018 results. X. Constraints on inflation},'' {\em Astron. Astrophys.}, vol.~641, p.~A10, 2020.

\bibitem{1990eaun.book.....K}
E.~W. {Kolb} and M.~S. {Turner}, {\em {The early universe}}, vol.~69.
\newblock 1990.

\bibitem{Tytler:2000qf}
D.~Tytler, J.~M. O'Meara, N.~Suzuki, and D.~Lubin, ``{Review of Big Bang nucleosynthesis and primordial abundances},'' {\em Phys. Scripta T}, vol.~85, p.~12, 2000.

\bibitem{deSalas2015glj}
P.~F. de~Salas, M.~Lattanzi, G.~Mangano, G.~Miele, S.~Pastor, and O.~Pisanti, ``{Bounds on very low reheating scenarios after Planck},'' {\em Phys. Rev. D}, vol.~92, no.~12, p.~123534, 2015.

\bibitem{Hasegawa2019jsa}
T.~Hasegawa, N.~Hiroshima, K.~Kohri, R.~S.~L. Hansen, T.~Tram, and S.~Hannestad, ``{MeV-scale reheating temperature and thermalization of oscillating neutrinos by radiative and hadronic decays of massive particles},'' {\em JCAP}, vol.~12, p.~012, 2019.

\bibitem{2021arXiv210410552E}
K.~{El Bourakadi}, ``{Preheating and Reheating after Standard Inflation},'' {\em arXiv e-prints}, p.~arXiv:2104.10552, Apr. 2021.

\bibitem{Giovannini:1998bp}
M.~Giovannini, ``{Gravitational waves constraints on postinflationary phases stiffer than radiation},'' {\em Phys. Rev. D}, vol.~58, p.~083504, 1998.

\bibitem{Giovannini:1999bh}
M.~Giovannini, ``{Production and detection of relic gravitons in quintessential inflationary models},'' {\em Phys. Rev. D}, vol.~60, p.~123511, 1999.

\bibitem{Domenech:2023jve}
G.~Dom\`enech, ``{Cosmological gravitational waves from isocurvature fluctuations},'' {\em AAPPS Bull.}, vol.~34, no.~1, p.~4, 2024.

\bibitem{Padilla:2023lbv}
L.~E. Padilla, J.~C. Hidalgo, and G.~German, ``{Constraining inflationary potentials with inflaton PBHs},'' {\em Phys. Rev. D}, vol.~108, no.~6, p.~063511, 2023.

\bibitem{2019NatAs...3..891V}
L.~{Verde}, T.~{Treu}, and A.~G. {Riess}, ``{Tensions between the early and late Universe},'' {\em Nature Astronomy}, vol.~3, pp.~891--895, Sept. 2019.

\bibitem{PhysRevLett.82.896}
I.~Zlatev, L.~Wang, and P.~J. Steinhardt, ``Quintessence, cosmic coincidence, and the cosmological constant,'' {\em Phys. Rev. Lett.}, vol.~82, pp.~896--899, Feb 1999.

\bibitem{doi:10.1146/annurev-astro-091916-055313}
J.~S. Bullock and M.~Boylan-Kolchin, ``Small-scale challenges to the lcdm paradigm,'' {\em Annual Review of Astronomy and Astrophysics}, vol.~55, no.~1, pp.~343--387, 2017.

\bibitem{PhysRevD261231}
A.~Vilenkin and L.~H. Ford, ``Gravitational effects upon cosmological phase transitions,'' {\em Phys. Rev. D}, vol.~26, Sep 1982.

\bibitem{Dine:1995uk}
M.~Dine, L.~Randall, and S.~D. Thomas, ``{Supersymmetry breaking in the early universe},'' {\em Phys. Rev. Lett.}, vol.~75, pp.~398--401, 1995.

\bibitem{Moroi:1999zb}
T.~Moroi and L.~Randall, ``{Wino cold dark matter from anomaly mediated SUSY breaking},'' {\em Nucl. Phys. B}, vol.~570, pp.~455--472, 2000.

\bibitem{Lyth:1995hj}
D.~H. Lyth and E.~D. Stewart, ``{Cosmology with a TeV mass GUT Higgs},'' {\em Phys. Rev. Lett.}, vol.~75, pp.~201--204, 1995.

\bibitem{Cicoli:2023opf}
M.~Cicoli, J.~P. Conlon, A.~Maharana, S.~Parameswaran, F.~Quevedo, and I.~Zavala, ``{String cosmology: From the early universe to today},'' {\em Phys. Rept.}, vol.~1059, pp.~1--155, 2024.

\bibitem{Asaka:1999xd}
T.~Asaka and M.~Kawasaki, ``{Cosmological moduli problem and thermal inflation models},'' {\em Phys. Rev. D}, vol.~60, p.~123509, 1999.

\bibitem{Dutta:2016htz}
B.~Dutta, E.~Jimenez, and I.~Zavala, ``{Dark Matter Relics and the Expansion Rate in Scalar-Tensor Theories},'' {\em JCAP}, vol.~06, p.~032, 2017.

\bibitem{TDamour_1992}
T.~Damour and G.~Esposito-Farese, ``Tensor-multi-scalar theories of gravitation,'' {\em Classical and Quantum Gravity}, vol.~9, p.~2093, sep 1992.

\bibitem{Cline:1999ts}
J.~M. Cline, C.~Grojean, and G.~Servant, ``{Cosmological expansion in the presence of extra dimensions},'' {\em Phys. Rev. Lett.}, vol.~83, p.~4245, 1999.

\bibitem{Csaki:1999jh}
C.~Csaki, M.~Graesser, C.~F. Kolda, and J.~Terning, ``{Cosmology of one extra dimension with localized gravity},'' {\em Phys. Lett. B}, vol.~462, pp.~34--40, 1999.

\bibitem{Josan:2009qn}
A.~S. Josan, A.~M. Green, and K.~A. Malik, ``{Generalised constraints on the curvature perturbation from primordial black holes},'' {\em Phys. Rev. D}, vol.~79, p.~103520, 2009.

\bibitem{Cole:2017gle}
P.~S. Cole and C.~T. Byrnes, ``{Extreme scenarios: the tightest possible constraints on the power spectrum due to primordial black holes},'' {\em JCAP}, vol.~02, p.~019, 2018.

\bibitem{Carr:2017edp}
B.~Carr, T.~Tenkanen, and V.~Vaskonen, ``{Primordial black holes from inflaton and spectator field perturbations in a matter-dominated era},'' {\em Phys. Rev. D}, vol.~96, no.~6, p.~063507, 2017.

\bibitem{Bhattacharya:2021wnk}
S.~Bhattacharya, A.~Das, and K.~Dutta, ``{Solar mass primordial black holes in moduli dominated universe},'' {\em JCAP}, vol.~10, p.~071, 2021.

\bibitem{Bhattacharya:2019bvk}
S.~Bhattacharya, S.~Mohanty, and P.~Parashari, ``{Primordial black holes and gravitational waves in nonstandard cosmologies},'' {\em Phys. Rev. D}, vol.~102, no.~4, p.~043522, 2020.

\bibitem{Harada:2016mhb}
T.~Harada, C.-M. Yoo, K.~Kohri, K.-i. Nakao, and S.~Jhingan, ``{Primordial black hole formation in the matter-dominated phase of the Universe},'' {\em Astrophys. J.}, vol.~833, no.~1, p.~61, 2016.

\bibitem{Harada:2017fjm}
T.~Harada, C.-M. Yoo, K.~Kohri, and K.-I. Nakao, ``{Spins of primordial black holes formed in the matter-dominated phase of the Universe},'' {\em Phys. Rev. D}, vol.~96, no.~8, p.~083517, 2017.
\newblock [Erratum: Phys.Rev.D 99, 069904 (2019)].

\bibitem{Kokubu:2018fxy}
T.~Kokubu, K.~Kyutoku, K.~Kohri, and T.~Harada, ``{Effect of Inhomogeneity on Primordial Black Hole Formation in the Matter Dominated Era},'' {\em Phys. Rev. D}, vol.~98, no.~12, p.~123024, 2018.

\bibitem{Passaglia:2021jla}
S.~Passaglia and M.~Sasaki, ``{Primordial black holes from CDM isocurvature perturbations},'' {\em Phys. Rev. D}, vol.~105, no.~10, p.~103530, 2022.

\bibitem{Green:2020jor}
A.~M. Green and B.~J. Kavanagh, ``{Primordial Black Holes as a dark matter candidate},'' {\em J. Phys. G}, vol.~48, no.~4, p.~043001, 2021.

\bibitem{ShamsEsHaghi:2022azq}
B.~Shams Es~Haghi, ``{Baryogenesis and primordial black hole dark matter from heavy metastable particles},'' {\em Phys. Rev. D}, vol.~107, no.~8, p.~083507, 2023.

\bibitem{Harada:2022xjp}
T.~Harada, K.~Kohri, M.~Sasaki, T.~Terada, and C.-M. Yoo, ``{Threshold of primordial black hole formation against velocity dispersion in matter-dominated era},'' {\em JCAP}, vol.~02, p.~038, 2023.

\bibitem{Bhattacharya:2023ztw}
S.~Bhattacharya, ``{Primordial Black Hole Formation in Non-Standard Post-Inflationary Epochs},'' {\em Galaxies}, vol.~11, no.~1, p.~35, 2023.

\bibitem{PBHBeta_documentation}
T.~D. Gomez-Aguilar and L.~E. Padilla, ``{PBHBeta: A Python Package for Calculating the Abundance of Primordial Black Holes}.'' Read the docs:~\url{https://pbhbeta.readthedocs.io/en}, 2023.

\bibitem{PBHBeta_repository}
T.~D. Gomez-Aguilar and L.~E. Padilla, ``{PBHBeta}.'' \url{https://github.com/TadeoDGAguilar/PBHBeta}, 2023.

\bibitem{2021OJAp....4E...1A}
R.~{Allahverdi}, M.~A. {Amin}, A.~{Berlin}, N.~{Bernal}, C.~T. {Byrnes}, M.~S. {Delos}, A.~L. {Erickcek}, M.~{Escudero}, D.~G. {Figueroa}, K.~{Freese}, T.~{Harada}, D.~{Hooper}, D.~I. {Kaiser}, T.~{Karwal}, K.~{Kohri}, G.~{Krnjaci}, M.~{Lewicki}, K.~D. {Lozanov}, V.~{Poulin}, K.~{Sinha}, T.~L. {Smith}, T.~{Takahashi}, T.~{Tenkanen}, J.~{Unwin}, and S.~{Watson}, ``{The First Three Seconds: a Review of Possible Expansion Histories of the Early Universe},'' {\em The Open Journal of Astrophysics}, vol.~4, p.~1, Jan. 2021.

\bibitem{2022arXiv221105767E}
A.~{Escriv{\`a}}, F.~{Kuhnel}, and Y.~{Tada}, ``{Primordial Black Holes},'' {\em arXiv e-prints}, p.~arXiv:2211.05767, Nov. 2022.

\bibitem{PhysRevLett.75.201}
D.~H. Lyth and E.~D. Stewart, ``Cosmology with a tev mass higgs field breaking the grand-unified-theory gauge symmetry,'' {\em Phys. Rev. Lett.}, vol.~75, pp.~201--204, Jul 1995.

\bibitem{PhysRevD.53.1784}
D.~H. Lyth and E.~D. Stewart, ``Thermal inflation and the moduli problem,'' {\em Phys. Rev. D}, vol.~53, pp.~1784--1798, Feb 1996.

\bibitem{Tsamis:2003px}
N.~C. Tsamis and R.~P. Woodard, ``{Improved estimates of cosmological perturbations},'' {\em Phys. Rev. D}, vol.~69, p.~084005, 2004.

\bibitem{Kinney:2005vj}
W.~H. Kinney, ``{Horizon crossing and inflation with large eta},'' {\em Phys. Rev. D}, vol.~72, p.~023515, 2005.

\bibitem{PhysRevD.28.1243}
M.~S. Turner, ``Coherent scalar-field oscillations in an expanding universe,'' {\em Phys. Rev. D}, vol.~28, pp.~1243--1247, Sep 1983.

\bibitem{MustafaA.Amin_2010}
M.~A. Amin, R.~Easther, and H.~Finkel, ``Inflaton fragmentation and oscillon formation in three dimensions,'' {\em Journal of Cosmology and Astroparticle Physics}, vol.~2010, p.~001, dec 2010.

\bibitem{PhysRevLett.108.241302}
M.~A. Amin, R.~Easther, H.~Finkel, R.~Flauger, and M.~P. Hertzberg, ``Oscillons after inflation,'' {\em Phys. Rev. Lett.}, vol.~108, p.~241302, Jun 2012.

\bibitem{PhysRevD.83.096010}
M.~Gleiser, N.~Graham, and N.~Stamatopoulos, ``Generation of coherent structures after cosmic inflation,'' {\em Phys. Rev. D}, vol.~83, p.~096010, May 2011.

\bibitem{2010arXiv1006.3075A}
M.~A. {Amin}, ``{Inflaton fragmentation: Emergence of pseudo-stable inflaton lumps (oscillons) after inflation},'' {\em arXiv e-prints}, p.~arXiv:1006.3075, June 2010.

\bibitem{Niemeyer:2019gab}
J.~C. Niemeyer and R.~Easther, ``{Inflaton clusters and inflaton stars},'' {\em JCAP}, vol.~07, p.~030, 2020.

\bibitem{PhysRevD.103.063525}
B.~Eggemeier, J.~C. Niemeyer, and R.~Easther, ``Formation of inflaton halos after inflation,'' {\em Phys. Rev. D}, vol.~103, p.~063525, Mar 2021.

\bibitem{Eggemeier:2021smj}
B.~Eggemeier, B.~Schwabe, J.~C. Niemeyer, and R.~Easther, ``{Gravitational collapse in the postinflationary Universe},'' {\em Phys. Rev. D}, vol.~105, no.~2, p.~023516, 2022.

\bibitem{Padilla:2021zgm}
L.~E. Padilla, J.~C. Hidalgo, and K.~A. Malik, ``{New mechanism for primordial black hole formation during reheating},'' {\em Phys. Rev. D}, vol.~106, no.~2, p.~023519, 2022.

\bibitem{Hidalgo:2022yed}
J.~C. Hidalgo, L.~E. Padilla, and G.~German, ``{Production of PBHs from inflaton structures},'' {\em {Phys. Rev. D}}, vol.~107, no.~6, p.~063519, 2023.

\bibitem{Padilla:2024iyr}
L.~E. Padilla, J.~C. Hidalgo, T.~D. Gomez-Aguilar, K.~A. Malik, and G.~German, ``{Primordial black hole formation during slow-reheating: a review},'' {\em Front. Astron. Space Sci.}, vol.~11, p.~1361399, 2024.

\bibitem{PhysRevD.97.023533}
K.~D. Lozanov and M.~A. Amin, ``Self-resonance after inflation: Oscillons, transients, and radiation domination,'' {\em Phys. Rev. D}, vol.~97, p.~023533, Jan 2018.

\bibitem{PhysRevLett.119.061301}
K.~D. Lozanov and M.~A. Amin, ``Equation of state and duration to radiation domination after inflation,'' {\em Phys. Rev. Lett.}, vol.~119, p.~061301, Aug 2017.

\bibitem{MOROI2000455}
T.~Moroi and L.~Randall, ``Wino cold dark matter from anomaly mediated susy breaking,'' {\em Nuclear Physics B}, vol.~570, no.~1, pp.~455--472, 2000.

\bibitem{PhysRevD.26.1231}
A.~Vilenkin and L.~H. Ford, ``Gravitational effects upon cosmological phase transitions,'' {\em Phys. Rev. D}, vol.~26, pp.~1231--1241, Sep 1982.

\bibitem{COUGHLAN198359}
G.~Coughlan, W.~Fischler, E.~W. Kolb, S.~Raby, and G.~Ross, ``Cosmological problems for the polonyi potential,'' {\em Physics Letters B}, vol.~131, no.~1, pp.~59--64, 1983.

\bibitem{PhysRevD.50.6357}
A.~A. Starobinsky and J.~Yokoyama, ``Equilibrium state of a self-interacting scalar field in the de sitter background,'' {\em Phys. Rev. D}, vol.~50, pp.~6357--6368, Nov 1994.

\bibitem{Kane:2015jia}
G.~Kane, K.~Sinha, and S.~Watson, ``{Cosmological Moduli and the Post-Inflationary Universe: A Critical Review},'' {\em Int. J. Mod. Phys. D}, vol.~24, no.~08, p.~1530022, 2015.

\bibitem{PhysRevD.80.083529}
B.~S. Acharya, G.~Kane, S.~Watson, and P.~Kumar, ``Nonthermal ``wimp miracle'','' {\em Phys. Rev. D}, vol.~80, p.~083529, Oct 2009.

\bibitem{Li:2016mmc}
B.~Li, P.~R. Shapiro, and T.~Rindler-Daller, ``{Bose-Einstein-condensed scalar field dark matter and the gravitational wave background from inflation: new cosmological constraints and its detectability by LIGO},'' {\em Phys. Rev. D}, vol.~96, no.~6, p.~063505, 2017.

\bibitem{Musco:2018rwt}
I.~Musco, ``{Threshold for primordial black holes: Dependence on the shape of the cosmological perturbations},'' {\em Phys. Rev. D}, vol.~100, no.~12, p.~123524, 2019.

\bibitem{Niemeyer:1997mt}
J.~C. Niemeyer and K.~Jedamzik, ``{Near-critical gravitational collapse and the initial mass function of primordial black holes},'' {\em Phys. Rev. Lett.}, vol.~80, pp.~5481--5484, 1998.

\bibitem{Musco:2008hv}
I.~Musco, J.~C. Miller, and A.~G. Polnarev, ``{Primordial black hole formation in the radiative era: Investigation of the critical nature of the collapse},'' {\em Class. Quant. Grav.}, vol.~26, p.~235001, 2009.

\bibitem{PhysRevD.101.044022}
A.~Escriv\`a, C.~Germani, and R.~K. Sheth, ``Universal threshold for primordial black hole formation,'' {\em Phys. Rev. D}, vol.~101, p.~044022, Feb 2020.

\bibitem{PhysRevD.103.063538}
I.~Musco, V.~De~Luca, G.~Franciolini, and A.~Riotto, ``Threshold for primordial black holes. ii. a simple analytic prescription,'' {\em Phys. Rev. D}, vol.~103, p.~063538, Mar 2021.

\bibitem{PhysRevD.81.104019}
B.~J. Carr, K.~Kohri, Y.~Sendouda, and J.~Yokoyama, ``New cosmological constraints on primordial black holes,'' {\em Phys. Rev. D}, vol.~81, p.~104019, May 2010.

\bibitem{Carr:2020gox}
B.~Carr, K.~Kohri, Y.~Sendouda, and J.~Yokoyama, ``{Constraints on primordial black holes},'' {\em Rept. Prog. Phys.}, vol.~84, no.~11, p.~116902, 2021.

\bibitem{Dalianis:2021dbs}
I.~Dalianis and G.~P. Kodaxis, ``{Reheating in Runaway Inflation Models via the Evaporation of Mini Primordial Black Holes},'' {\em Galaxies}, vol.~10, no.~1, p.~31, 2022.

\bibitem{Martin:2019nuw}
J.~Martin, T.~Papanikolaou, and V.~Vennin, ``{Primordial black holes from the preheating instability in single-field inflation},'' {\em JCAP}, vol.~01, p.~024, 2020.

\bibitem{Keith:2020jww}
C.~Keith, D.~Hooper, N.~Blinov, and S.~D. McDermott, ``{Constraints on Primordial Black Holes From Big Bang Nucleosynthesis Revisited},'' {\em Phys. Rev. D}, vol.~102, no.~10, p.~103512, 2020.

\bibitem{PhysRevD.41.3052}
J.~H. MacGibbon and B.~R. Webber, ``Quark- and gluon-jet emission from primordial black holes: The instantaneous spectra,'' {\em Phys. Rev. D}, vol.~41, pp.~3052--3079, May 1990.

\bibitem{MacGibbon:1991vc}
J.~H. MacGibbon and B.~J. Carr, ``{Cosmic rays from primordial black holes},'' {\em Astrophys. J.}, vol.~371, pp.~447--469, 1991.

\bibitem{Starobinsky:1973aij}
A.~A. Starobinsky, ``{Amplification of waves reflected from a rotating ''black hole''.},'' {\em Sov. Phys. JETP}, vol.~37, no.~1, pp.~28--32, 1973.

\bibitem{Clark:2016nst}
S.~Clark, B.~Dutta, Y.~Gao, L.~E. Strigari, and S.~Watson, ``{Planck Constraint on Relic Primordial Black Holes},'' {\em Phys. Rev. D}, vol.~95, no.~8, p.~083006, 2017.

\bibitem{LEMOINE2000333}
M.~Lemoine, ``Moduli constraints on primordial black holes,'' {\em Phys. Lett. B}, vol.~481, no.~2, pp.~333--338, 2000.

\bibitem{MacGibbon:1987my}
J.~H. MacGibbon, ``{Can Planck-mass relics of evaporating black holes close the universe?},'' {\em Nature}, vol.~329, pp.~308--309, 1987.

\bibitem{2007hep.th....3070A}
S.~{Alexander} and P.~{M{\'e}sz{\'a}ros}, ``{Reheating, Dark Matter and Baryon Asymmetry: a Triple Coincidence in Inflationary Models},'' {\em arXiv e-prints}, pp.~hep--th/0703070, Mar. 2007.

\bibitem{Press:1973iz}
W.~H. Press and P.~Schechter, ``{Formation of galaxies and clusters of galaxies by selfsimilar gravitational condensation},'' {\em Astrophys. J.}, vol.~187, pp.~425--438, 1974.

\bibitem{Carr:1975qj}
B.~J. Carr, ``{The Primordial black hole mass spectrum},'' {\em Astrophys. J.}, vol.~201, pp.~1--19, 1975.

\bibitem{Niemeyer:1999ak}
J.~C. Niemeyer and K.~Jedamzik, ``{Dynamics of primordial black hole formation},'' {\em Phys. Rev. D}, vol.~59, p.~124013, 1999.

\bibitem{Hawke:2002rf}
I.~Hawke and J.~M. Stewart, ``{The dynamics of primordial black hole formation},'' {\em Class. Quant. Grav.}, vol.~19, pp.~3687--3707, 2002.

\bibitem{Bardeen:1985tr}
J.~M. Bardeen, J.~R. Bond, N.~Kaiser, and A.~S. Szalay, ``{The Statistics of Peaks of Gaussian Random Fields},'' {\em Astrophys. J.}, vol.~304, pp.~15--61, 1986.

\bibitem{2024arXiv240200600A}
P.~{Auclair} and B.~{Blachier}, ``{Small-scale clustering of Primordial Black Holes: cloud-in-cloud and exclusion effects},'' {\em arXiv e-prints}, p.~arXiv:2402.00600, Feb. 2024.

\bibitem{Erfani:2021rmw}
E.~Erfani, H.~Kameli, and S.~Baghram, ``{Primordial black holes in the excursion set theory},'' {\em Mon. Not. Roy. Astron. Soc.}, vol.~505, no.~2, pp.~1787--1793, 2021.

\bibitem{Schunck:2003kk}
F.~E. Schunck and E.~W. Mielke, ``{General relativistic boson stars},'' {\em Class. Quant. Grav.}, vol.~20, pp.~R301--R356, 2003.

\bibitem{Dawoodbhoy:2021beb}
T.~Dawoodbhoy, P.~R. Shapiro, and T.~Rindler-Daller, ``{Core-envelope haloes in scalar field dark matter with repulsive self-interaction: fluid dynamics beyond the de Broglie wavelength},'' {\em Mon. Not. Roy. Astron. Soc.}, vol.~506, no.~2, pp.~2418--2444, 2021.

\bibitem{DeLuca:2021pls}
V.~De~Luca, G.~Franciolini, A.~Kehagias, P.~Pani, and A.~Riotto, ``{Primordial black holes in matter-dominated eras: The role of accretion},'' {\em Phys. Lett. B}, vol.~832, p.~137265, 2022.

\bibitem{Burgess:2010bz}
C.~P. Burgess, M.~Cicoli, M.~Gomez-Reino, F.~Quevedo, G.~Tasinato, and I.~Zavala, ``{Non-standard primordial fluctuations and nongaussianity in string inflation},'' {\em JHEP}, vol.~08, p.~045, 2010.

\bibitem{Domenech:2019quo}
G.~Dom\`enech, ``{Induced gravitational waves in a general cosmological background},'' {\em Int. J. Mod. Phys. D}, vol.~29, no.~03, p.~2050028, 2020.

\bibitem{Domenech:2020kqm}
G.~Dom\`enech, S.~Pi, and M.~Sasaki, ``{Induced gravitational waves as a probe of thermal history of the universe},'' {\em JCAP}, vol.~08, p.~017, 2020.

\bibitem{Domenech:2021ztg}
G.~Dom\`enech, ``{Scalar Induced Gravitational Waves Review},'' {\em Universe}, vol.~7, no.~11, p.~398, 2021.

\bibitem{Papanikolaou:2020qtd}
T.~Papanikolaou, V.~Vennin, and D.~Langlois, ``{Gravitational waves from a universe filled with primordial black holes},'' {\em JCAP}, vol.~03, p.~053, 2021.

\bibitem{Papanikolaou:2022chm}
T.~Papanikolaou, ``{Gravitational waves induced from primordial black hole fluctuations: the~effect of an extended mass function},'' {\em JCAP}, vol.~10, p.~089, 2022.

\bibitem{Germani:2019zez}
C.~Germani and R.~K. Sheth, ``{Nonlinear statistics of primordial black holes from Gaussian curvature perturbations},'' {\em Phys. Rev. D}, vol.~101, no.~6, p.~063520, 2020.

\end{thebibliography}

\end{document}